\documentclass[twocolumn,secnumarabic,amssymb, nobibnotes, aps, prd]{revtex4}
\usepackage{graphicx}
\usepackage{longtable}

\begin{document}

\title{Energy-momentum Prescriptions in General Spherically Symmetric Space-times}%

\author{Saeed Mirshekari}
\email{smirshekari@ut.ac.ir}
\author{Amir M. Abbassi}
\email{amabasi@khayam.ut.ac.ir} \affiliation{Department of
Physics, University of Tehran, North Kargar Ave,\\ Tehran, Iran.}
\date{July, 2008}
\begin{abstract}
Einstein, Landau-Lifshitz, Papapetrou, Weinberg, and M\o ller
energy-momentum prescriptions in general spherically symmetric
space-times are investigated. It is shown that for two special but
not unusual classes of general spherically symmetric space-times
several energy-momentum prescriptions in Schwarzschild Cartesian
coordinates lead to some coincidences in energy distribution. It
is also obtained that for a special class of spherically symmetric
metrics M\o ller and Einstein energy-momentum prescriptions give
the same result for energy distribution if and only if it has a
specific dependence on radial coordinate.

\keywords{energy-momentum prescriptions, spherically symmetric
space-time.}
\pacs{04.20.-q, 04.20.Cv, 98.80.Cq,
04.20.Jb,98.80.Es}
\end{abstract}

\maketitle

\section{Introduction}
One of the old and basic problems in General Relativity (GR) which
is still unsolved is the localization of energy. In Special
Relativity one can define a symmetric tensor, $T_a^b$, as
energy-momentum tensor which satisfies the conservation laws, i.e.
\begin{eqnarray}\label{r1}
T^b_{a ,b}=0,
\end{eqnarray}
that means the energy-momentum tensor is a conserved and localized
quantity in special relativity (and classical mechanics). In fact,
in any local point of manifold no contribution of this quantity is
produced or eliminated. But, in GR Eq.(\ref{r1}) is not
acceptable. Because, it is not a tensor equation and is not valid
in all reference frames. Using covariant derivative instead of
partial one, we reach to the following equation which is invariant
in all reference frames \cite{weinberg}.
\begin{eqnarray}\label{r2}
T^b_{a
;b}=\frac{1}{\sqrt{-g}}(\sqrt{-g}T^b_{a})_{,b}-\Gamma^b_{ac}T_b^c=0
\end{eqnarray}
where $\Gamma^b_{ac}$ are the connection coefficients. Considering
Eq.(\ref{r2}), it is obvious that Eq.(\ref{r1}) is no longer
satisfied. So, in this situation this energy-momentum tensor is
not a localized quantity and the problem of energy-momentum
localization in GR arises. If we want to keep the localization
characteristics of energy in GR, we must look for a new quantity
such as $_{\textrm{\tiny eff}}T_a^b$, instead of $T_a^b$, which
its partial derivative vanishes, i.e.
\begin{eqnarray}
_{\textrm{\tiny eff}}T_{a ,b}^b=0.
\end{eqnarray}

Considering the relation between partial and covariant derivatives
Eq.(\ref{r2}), the following would be a suitable candidate
\cite{misner}

\begin{equation} \label{3}
_{\textrm{\tiny
eff}}T^{b}_{a}=(-g)^{\frac{n}{2}}(T^{b}_{a}+t^{b}_{a})
\end{equation}
where $g=det(g_{ab})$ and $n$ is a positive integer that shows the
weight. In fact, in this way we consider gravitational fields
effects in energy distribution as an additional term in
energy-momentum tensor. In other words, $_{\textrm{\tiny
eff}}T^{b}_{a}$ is an energy-momentum complex of matter plus
gravitational fields where $t^{b}_{a}$ is not a true tensor, but
rather is a pseudo-tensor that describes the localization of
gravitational energy-momentum.

\medskip

It should be noted that $ _{\textrm{\tiny eff}}T_{a}^{b} $  can be
written as the divergence of some ``super-potential'' $ H_{a}^{[b
c]} $  that is anti-symmetric in its two upper indices
\cite{favata} as
\begin{equation}
_{\textrm{\tiny eff}}T_{a}^{b}=H_{a ,c}^{[b c]}.
\end{equation}
In addition, a new quantity like $ U_{a}^{b c} $ can also play the
role of $ H_{a}^{[b c]} $ if
\begin{eqnarray}
U_{a}^{b c}=H_{a}^{[b c]}+\Psi_{a}^{b c}
\end{eqnarray}
and divergence or double divergence of $\Psi_{a}^{b c}$ is
identically zero, i.e.
\begin{eqnarray}
\Psi_{a ,c}^{b c}\equiv 0 \:\: or \:\: \Psi_{a ,cb}^{b c} \equiv
0.
\end{eqnarray}

So, the quantity $ \Theta_{a}^{b} $  which is defined by this new
super-potential remains conserved locally as
\begin{equation}
\Theta_{a}^{b}=U_{a ,c}^{b c}\Rightarrow \Theta_{a ,b}^{b}=0.
\end{equation}

Considering this freedom on the choice of superpotentials, many
different energy-momentum prescriptions (EMPs) have been proposed
by different authors, for example Einstein \cite{moller}, Landau
and Lifshitz \cite{landau}, M\o ller \cite{moller}, Bergmann
\cite{bergmann}, Weinberg \cite{weinberg}, Papapetrou
\cite{papapetrou}, Tolman$^1$\footnotetext[1]{Although the Tolman
and Einstein prescriptions are different in their forms, but in
fact they are equal \cite{virbb}.} \cite{tolman}, Komar
\cite{komar}, Penrose \cite{penrose} and Qadir and Sharif
\cite{qadir} prescriptions.

\medskip

Using EMPs has some problems that are mentioned in the following.
Except a few of them including M\o ller, Penrose, and Komar
prescriptions, for other prescriptions all calculations must be
done in Cartesian coordinate system. Moreover, some of them are
non-symmetric in exchanging of their indices. So, conserved
angular momentum can not be defined for that ones which are
non-symmetric. Another drawback of using EMPs is that they may
give different results for the same space-time. Finally, physical
concept of these non-tensorial quantities has been unclear for a
long time. However, Cheng, Nester, and Chen \cite{chang} showed
that they can be considered as the boundary term of Hamiltonian
and therefore are quasi-local.

\medskip

The problems associated with the concept of energy-momentum
complexes resulted in some researchers even doubting the concept
of energy-momentum localization. Misner et al. \cite{misner}
argued that to look for a local energy-momentum is looking for the
right answer to the wrong question. He showed that the energy can
be localized only in systems which have spherical symmetry. They
also expressed that pseudo-tensor approach could conflict with the
equivalence principle. Cooperstock and Sarracino
\cite{cooperstock} argued that if energy is localizable for
spherical systems, then it can be localized in any system. In
1990, Bondi \cite{bondi} argued that a non-localizable form of
energy is not allowed in GR. Recently, besides EMPs, it was
suggested another viewpoint for energy problem in GR that is in
agreement with EMPs theory about localization of energy, i.e. EMPs
in ``Tele-Parallel Gravity'', (see \cite{teleparalel}). On the
other hand, some people do not believe in localization of energy
and momentum in GR. In addition, some physicists propose a new
concept in this regard: ``quasi-localization''. A large number of
definitions of quasi-local masses have been proposed
\cite{quasilocal}. Unlike EMPs theory, quasi-localization theory
does not restrict one to use particular coordinate system, but
this theory have also its drawbacks (see Bergqvist \cite{berq}).
In general, there is no generally accepted definition for energy
and momentum in GR till now.

\medskip

Despite of mentioned drawbacks of using EMPs, many authors have
been interested in this topic and have reached interesting results
\cite{virbhadra1, radinschi1, examples, virbb, aguirregabiria,
xulu, einmol, vagenas1, examples2}. Especially Virbhadra and
colleagues \cite{virbb, aguirregabiria} showed that for a specific
class of space-times, i.e. Kerr-Schild class, and even more
general space-times, Einstein, Landau-Lifshitz, Papapetrou, and
Weinberg prescriptions (ELLPW) lead to the same results in
KerrSchild Cartesian coordinates. Moreover, Virbhadra \cite{virbb}
and Xulu \cite{xulu} used ELLPW prescriptions and M\o ller
prescription in general spherically symmetric (GSS) space-time,
respectively, and found different expressions for the energy in a
sphere with radius $r$ in Schwarzschild Cartesian coordinates.
But, in Kerr-Schild Cartesian coordinates ELLPW prescriptions lead
to the same results in GSS space-time. It is not clear why
different EMPs ``coincide'' in the Kerr-Schild Cartesian
coordinates, but not in the Schwarzschild Cartesian coordinates.
Considering this confusion, in this paper we focus on EMPs in GSS
space-times in Schwarzschild Cartesian coordinates.

\medskip

There is no proved performance between different EMPs. However,
Palmer \cite{palmer} and Virbhadra \cite{virbb} discussed the
importance of Einstein EMP while Lessner \cite{lessner} believed
that M\o ller prescription is a powerful tool for calculating the
energy-momentum in GR. Recently, some authors have presented their
interest to study M\o ller and Einstein prescriptions and to find
a relation between them \cite{xulu, einmol, vagenas1}. They have
tried to compare these prescriptions in different space-times.

\medskip

The remainder of the paper is organized as follows. In section 2
we introduce several EMPs which are used in this paper and give
previous obtained results for these prescriptions in GSS
space-time in Schwarzschild Cartesian coordinates. In section 3 we
consider two special but not unusual spherically symmetric
space-times and calculate their energy distribution by using
different EMPs. In section 4, considering mentioned classes of GSS
metrics, we find a unique form of this class in which Einstein and
M\o ller prescriptions lead to the same result. Finally, we
summarize and conclude in section 5.

\medskip

\textit{Conventions:} We use geometrized units in which $
c=G=\hbar=1 $ and the metric has signature $(+ - - -)$. Latin
indices take values 0...3. The comma and semicolon, respectively,
stand for the partial and covariant derivatives.

\section{Energy Distribution of The Most General Non-static Spherically Symmetric Space-time }

\subsection{Energy-momentum Prescriptions (EMPs)}

In this section we introduce several EMPs which are used in this
paper, i.e. Einstein, Landau-Lifshitz, Papapetrou, Weinberg, and
M\o ller prescriptions. Specific form of each energy-momentum
pseudo-tensor, conservation laws, and energy-momentum 4-vectors
are listed in Table \ref{tableI}. Weinberg, Landau-Lifshitz, and
Papapetrou pseudo-tensors are symmetric in exchanging their
indices and so, using them, one can define a conserved angular
momentum. Moreover, we must perform our calculations in Cartesian
coordinate system in all of above-mentioned prescriptions, except
M\o ller prescription in which all coordinate systems are
acceptable. Interested readers can refer to the mentioned
references for more details. As mentioned in the previous section,
prescriptions's differences are just in a curl term. This topic
was discussed in reference \cite{favata} in more detail.

\medskip

In the third column of Table \ref{tableI}, Gauss' theorem is used.
In the surface integrals $ n_{a} $ represents the components of a
normal one form over an infinitesimal surface element $ds$. In
spherically symmetric space-times, suitable surface of integration
would be a sphere with radius $r$. In addition, $ds$ would be
equal to $r^2\sin(\theta)d\theta d\phi$. The results of
calculations according to each of the individual forms in GSS
space-times will be presented in the following subsections.

\begin{table*}
  \centering
  \caption{Energy-momentum Prescriptions}\label{tableI}

\begin{ruledtabular}
\begin{tabular}{|c|c|c|c|}
&&&\\
Prescription& Energy-momentum Pseudo-tensor&Conservation&Energy-momentum\\
&& Laws&4-vector\\
&&&\\
\hline
&&&\\
  Einstein & $\Theta_{i}^{k}=\frac{1}{16\pi}H_{i ,l}^{k l}$
& $\frac{\partial \Theta_{i} ^{k} }{\partial x^{k}}=0$
 & $P_{i}=\int\int\int \Theta_{i} ^{0}dx^{1}dx^{2}dx^{3}$ \\
$[3]$&$H_{i}^{kl}=-H_{i}^{lk}=\frac{g_{in}}{\sqrt{-g}}[-g(g^{kn}g^{lm}-g^{ln}g^{km})]_{,m}$&&$=\frac{1}{16\pi}\int\int H_{i}^{0 a}n_{a}ds$\\
&&&\\
\hline
&&&\\
  Landau-Lifshitz&$ L^{ik}=\frac{1}{16\pi}\lambda^{iklm}_{,lm}$& $\frac{\partial L^{i k} }{\partial x^{k}}=0
 $& $P^{i}=\int\int\int L^{i 0}dx^{1}dx^{2}dx^{3}$\\
$[4]$& $\lambda^{iklm}=-g(g^{ik}g^{lm}-g^{il}g^{km})_{,m}$& &$=\frac{1}{16\pi}\int\int \lambda^{i0\alpha m}_{,m}n_{a}ds$\\
&&&\\
\hline
&&&\\
& $\Sigma^{ik}=\frac{1}{16\pi}N^{iklm}_{,lm}$&&$ P^{i}=\int\int\int \Sigma^{i 0}dx^{1}dx^{2}dx^{3}$\\

 Papapetrou &$N^{iklm}=\sqrt{-g}(g^{ik}\eta^{lm}-g^{il}\eta^{km}+g^{lm}\eta^{ik}-g^{lk}\eta^{im})$&$\frac{\partial \Sigma^{i k} }{\partial x^{k}}=0$&$=\frac{1}{16\pi}\int\int N^{i0\alpha m}_{,m}n_{a}ds$\\

$[5]$ &$\eta^{ik}=diag(1,-1,-1,-1)$&&\\
&&&\\
\hline
&&&\\
&$W^{ik}=\frac{1}{16\pi}D^{i k l}_{,l}$&&$P^{i}=\int\int\int
W^{i0}dx^{1}dx^{2}dx^{3}$
\\

Weinberg&$D^{ijk}=\frac{\partial{h_{a}^{a}}}{\partial{x_{i}}}\eta^{jk}-
\frac{\partial{h_{a}^{a}}}{\partial{x_{j}}}\eta^{ik}-\frac{\partial{h^{ai}}}{\partial{x^{a}}}\eta^{jk}+\frac{\partial{h^{aj}}}{\partial{x^{a}}}\eta^{ik}+\frac{\partial{h^{ik}}}{\partial{x_{j}}}-\frac{\partial{h^{jk}}}{\partial{x_{i}}}$&$\frac{\partial
W ^{ik} }{\partial x^{k}}=0$&$=\frac{1}{16\pi}\int\int
D^{i0a}n_{a}ds
$\\
$[1]$&$h_{ik}=g_{ik}-\eta_{ik}$&&\\
&&&\\
\hline
&&&\\
M\o ller & $M_{i}^{k}=\frac{1}{8\pi}\chi_{i ,l}^{k l}$ &$
\frac{\partial M_{i} ^{k} }{\partial x^{k}}=0$ &$
P_{i}=\int\int\int M_{i} ^{0}dx^{1}dx^{2}dx^{3}$\\
$[3]$& $\chi_{i}^{k l}=\sqrt{-g}(\frac{\partial g_{ip}}{\partial
x^{q}}-\frac{\partial g_{iq}}{\partial x^{p}})g^{kq}g^{lp}$&
&$=\frac{1}{8\pi}\int\int\chi_{i}^{0 a}n_{a}ds$\\
&&&\\

\end{tabular}
\end{ruledtabular}
\end{table*}

\bigskip

\subsection{General Spherically Symmetric (GSS) Space-time}
Most general non-static spherically symmetric space-time is
described by the following line element.

\begin{eqnarray}\label{GSS}
ds^2&=&B(r,t) dt^2-A(r,t)dr^2\\\nonumber &&-2F(r,t) dt dr- D(r,t)
r^2 (d\theta^2+\sin^2\theta d\phi^2).
\end{eqnarray}

Transforming the line element (\ref{GSS}) which is in spherical
coordinates $(t, r, \theta, \phi)$ into Cartesian coordinates $(t,
x, y, z)$, we have (according to $x=r\sin\theta \cos\phi$,
$y=r\sin\theta \sin\phi$, $z=r\cos\theta$)

\begin{eqnarray}\label{carGSS}
ds^2&=&B(r,t) dt^2-2 F(r,t)\nonumber
(\frac{x}{r}dx+\frac{y}{r}dy+\frac{z}{r}dz) dt\\
&&-D(r,t)(dx^2+dy^2+dz^2)\\\nonumber &&(D(r,t)-A(r,t))(\frac{x dx
+y dy+ z dz}{r})^2.\\\nonumber
\end{eqnarray}

Using Einstein, Landau-Lifshitz, and Weinberg prescriptions for
the metric (\ref{carGSS}), the energy for each EMP inside a
2-sphere with radius $r$ is given by \cite{virbb}

\begin{eqnarray}\label{Einstein}
_{\textrm{\tiny E}}E&=&\frac{r}{2}\frac{B(A-D-D' r)-F(r
\dot{D}-F)}{2\sqrt{A B}+F^2},\\\label{Landau} _{\textrm{\tiny
LL}}E&=&\frac{r}{2}\frac{D(A-D-D' r)}{2},\\\label{Weinberg}
_{\textrm{\tiny W}}E&=&\frac{r}{2}\frac{(A-D-D' r)}{2}
\end{eqnarray}
where prime and dot denote the partial derivatives with respect to
the coordinates $r$ and $t$, respectively. For static and
spherically symmetric space-times ($F=0$; $A$, $B$, and $D$ only
depend on $r$ coordinate) we have \cite{virbb}

\begin{eqnarray}\label{Papetrou}
_{\textrm{\tiny P}}E&=&\frac{r}{8 (A B)^\frac{3}{2}}[4 A
B^2\nonumber (A-D)+ r (A^2 B' D\\ &&- 2 A^2 B D'- A A' B D- 2 A
B^2 D'\\\nonumber &&- A B B' D+ A' B^2 D)].\\\nonumber
\end{eqnarray}

Using M\o ller prescription for line element (\ref{GSS}) or
(\ref{carGSS}) one can find \cite{xulu}

\begin{eqnarray}\label{Moller}
_{\textrm{\tiny M}}E&=&\frac{r^2}{2}\frac{D (B'+\dot{F})}{\sqrt{A
B+ F^2}}
\end{eqnarray}

It is obvious from Eqs.(\ref{Einstein}, \ref{Landau},
\ref{Weinberg}, \ref{Moller}) that EMPs disagree for the most
general non-static spherically symmetric space-time. It should be
noted that in obtaining all above expressions for the energy,
Schwarzschild Cartesian coordinate ($t, r, \theta, \phi$) have
been used, while using Kerr-Schild Cartesian coordinates for a
general non-static spherically symmetric space-time of the
Kerr-Schild class described with the line element
\begin{eqnarray}\label{kerrschild}
ds^2=B(u,r) du^2- 2 du dr- r^2 (d\theta^2+ \sin\theta^2 d\phi^2)
\end{eqnarray}
leads to the same result for Einstein, Landau-Lifshitz,
Papapetrou, and Weinberg prescriptions given by \cite{virbb}
\begin{eqnarray}
E&=&\frac{r}{2}(1-B(u)).
\end{eqnarray}
In Eq.(\ref{kerrschild}), $u$ coordinate is related to both $t$
and $r$ coordinates as $u=r+t$.

\medskip

According to Virbhadra in reference \cite{virbb} \textit{``It is
not clear why different definitions coincide when calculations are
carried out in Kerr-Schild Cartesian coordinates, but disagree in
Schwarzschild Cartesian coordinates. At this stage it is not known
if this is accidental or points out something interesting.''}.
This viewpoint is our main motivation to study some special cases
of the line element (\ref{GSS}) and to investigate EMPs in
Schwarzschild Cartesian coordinate if there is any coincidence
between them in these cases.

\bigskip

\section{Special Cases of GSS}
Considering Eq.(\ref{Landau}) and Eq.(\ref{Weinberg}), it is
obvious that Landau-Lifshitz and Weinberg prescriptions give the
same result if and only if $D=1$. Moreover, if $D=1$, $F=0$, and
$B=A^{-1}$ we have $_{\textrm{\tiny E}}E=\frac{r}{2}(1-B)$. In
addition, if $F=0$ and $D=1$ and we want to reach to the same
results for Einstein, Landau-Lifshitz, and Weinberg prescriptions
we should have $A+B=2$. In addition to $F=0$ and $D=1$ if
$B=A^{-1}$, then for coincidence of Einstein, Landau-Lifshitz, and
Weinberg prescriptions we should have $B=A=1$ (flat space-time).

\medskip

In the following subsections we consider two special but not
unusual spherically symmetric space-times which lead to
interesting results for energy distribution. In the first case we
assume $D=1$, $F=0$, $B=A^{-1}=(1-f)^\mu$ and in the next one it
is supposed that $D=(1-f)^{1-\mu}$, $F=0$, and
$B=A^{-1}=(1-f)^\mu$ where $f$ is only a function of $r$, i.e.
$f(r)$, and $\mu$ is a constant number. Some examples of the first
case are listed in Table \ref{tableII}. As one can see in this
Table, many important and well-known metrics belong to the above
mentioned special case. In the two following subsections, using
different EMPs we calculate the energy expressions and compare
them with each other.

\begin{table}
\centering \caption{Some examples of a special case of GSS, i.e.
$D=1$, $F=0$, $B=A^{-1}=(1-f)^\mu$ in
Eq.(\ref{GSS})}\label{tableII}
\begin{ruledtabular}
\begin{tabular}{|c|c|c|}
&&\\
Metric & $f(r)$ & $\mu$ \\
&&\\
\hline
&&\\
Schwarzschild & $\frac{2m}{r}$&1\\
&&\\
\hline
&&\\
Reissner-Nordestr\"{o}m & $\frac{2m}{r}-\frac{q^2}{r^2}$&1\\
&&\\
\hline
&&\\
de-Sitter & $\frac{\Lambda}{3}r^2$&1 \\
&&\\
\hline
&&\\
Schwarzschild de-Sitter & $\frac{2m}{r}+\frac{\Lambda}{3}r^2$&1 \\
&&\\
\hline
&&\\
RN (anti)de-Sitter & $\frac{2m}{r}-\frac{q^2}{r^2}+\frac{|\Lambda|}{3}r^2$&1 \\
&&\\
\hline
&&\\
Dymnikova \cite{dym} & $\frac{R_g(r)}{r}$&1\\
&$R_g(r)=r_g(1-e^{\frac{r^3}{r_*^3}})$& \\
&&\\
\hline
&&\\
ABG Black Hole \cite{abg} & $\frac{2m}{r}(1-\tanh\frac{q^2}{2mr})$&1 \\
&&\\
\hline
&&\\
Conformal Scalar Dyon Black Hole & $\frac{Q_{\tiny{CSD}}}{r}$&2 \\
\cite{con}&&\\
\hline
&&\\
De Lorenci \cite{del} & $\frac{2m}{r}-\frac{q^2}{r^2}+\frac{\sigma Q^4}{5r^6}$&1 \\
&&\\
\hline
&&\\
Charged Topological Black Hole& $1+\frac{\Lambda}{3}r^2+(1-\frac{Gm}{r^2})^2$&1 \\
\cite{cha}&&\\
\hline
&&\\
Bardeen's Regular Black Hole \cite{bar}& $\frac{2m r^2}{(r^2+q^2)^{\frac{3}{2}}}$&1 \\
&&\\
\end{tabular}
\end{ruledtabular}
\end{table}

\medskip

\subsection{Case I, $D=1$}
Substituting $D=1$, $F=0$, $B=A^{-1}=(1-f)^\mu$ in the line
element (\ref{carGSS}) and considering Eqs.(\ref{Einstein},
\ref{Landau}, \ref{Weinberg}, \ref{Papetrou}, \ref{Moller}) we
find

\begin{eqnarray}\label{case1E}
_{\textrm{\tiny E}}E&=&-\frac{r}{2}((1-f)^\mu-1)\\\label{case1LLW}
_{\textrm{\tiny LL}}E&=&_{\textrm{\tiny
W}}E=\frac{r}{2}((1-f)^{-\mu}-1)\\\label{case1LLP} _{\textrm{\tiny
P}}E&=&-\frac{r}{4}[-2+ 2(1-f)^\mu +r(1-f)^{-\mu-1} \mu
f'\\\nonumber && -r(1-f)^{\mu-1} \mu f']\\\label{case1M}
_{\textrm{\tiny M}}E&=&-\frac{r^2}{2}(1-f)^{\mu-1} \mu f'.
\end{eqnarray}
Substituting mentioned conditions in case I ($D=1$, $F=0$,
$B=A^{-1}=(1-f)^\mu$) in Eq.(\ref{GSS}), it is not difficult to
find that $f$ indicates a perturbation term from flat space-time.
For small values of $f$ ($f\ll1$), Eqs.(\ref{case1E},
\ref{case1LLW}, \ref{case1LLP}, \ref{case1M}) reduce to
\begin{eqnarray}\label{c1}
_{\textrm{\tiny E}}E=_{\textrm{\tiny W}}E=_{\textrm{\tiny
LL}}E&=&\frac{r}{2}\mu f\\\label{c2} \label{d1papapetrou}
_{\textrm{\tiny P}}E&=&\frac{r}{2}\mu f-\frac{r^2}{2}\mu^2
ff'\\\label{c3} \label{d1moller} _{\textrm{\tiny
M}}E&=&-\frac{r^2}{2}(1-(\mu-1)f)\mu f'
\end{eqnarray}
The aforementioned equations are in full agreement with previous
results about energy distribution of some specific spherically
symmetric space-times with described conditions (see
\cite{virbhadra1, radinschi1, virbb, xulu, vagenas1, examples2}).

\medskip

It is interesting to note that when $r\rightarrow \infty$ all of
above expressions will be equal if
\begin{eqnarray}\label{r25}
f=\sum_{n} C_n r^{-n}
\end{eqnarray}
where $C_n$ are constant and $n$ is a positive integer number.
Recent condition means that this metric belongs to asymptotically
flat space-times. Considering Table \ref{tableII}, one can find
that Schwarzschild, Reissner-Nordestr\"{o}m, ABG Black Hole,
Conformal Scalar Dyon Black Hole, De Lorenci, Charged Topological
Black Hole, and Bardeen's Regular Black Hole space-times in this
table satisfy Eq.(\ref{r25}). In this situation only first terms
of Eqs.(\ref{c1}, \ref{c2}, \ref{c3}) remain and next terms
disappear. So, we obtain the following equivalence between the
results

\begin{eqnarray}
_{\textrm{\tiny E}}E=_{\textrm{\tiny LL}}E=_{\textrm{\tiny
P}}E=_{\textrm{\tiny W}}E=_{\textrm{\tiny M}}E&=&\frac{r}{2}\mu f.
\end{eqnarray}

\medskip

\subsection{Case II, $D=(1-f)^{1-\mu}$}
Assuming $D=(1-f)^{1-\mu}$, $F=0$, $B=A^{-1}=(1-f)^\mu$,
considering Eqs.(\ref{Einstein}, \ref{Landau}, \ref{Weinberg},
\ref{Papetrou}, \ref{Moller}), and using them in the line element
(\ref{GSS}), we find

\begin{eqnarray}
_{\textrm{\tiny E}}E&=&-\frac{r}{2}(-f- r f'+ r \mu f')\\
_{\textrm{\tiny LL}}E&=&\frac{r}{2}(1-f)^{-2\mu}\\&&\nonumber\times(f-f^2+ r f'(1-f-\mu+\mu f))\\
_{\textrm{\tiny W}}E&=&-\frac{r}{2}((1-f)^{-\mu}(-f-rf'+r\mu f'))\\
_{\textrm{\tiny P}}E&=&-\frac{r}{4}(-2f-r f'-(1-f)^{-2\mu}(r f'-2f'))\\
_{\textrm{\tiny M}}E&=&-\frac{r^2}{2}\mu f'
\end{eqnarray}
and for $f\ll1$ we have
\begin{eqnarray}
_{\textrm{\tiny E}}E=_{\textrm{\tiny W}}E=_{\textrm{\tiny
P}}E&=&-\frac{r}{2}(-f-r f'+ r\mu f').
\end{eqnarray}

It should be noted that for well-known Janis-Newman-Winicour (JNW)
space-time \cite{JNW} in which $f=\frac{2 M_*}{r}$,
$\mu=\frac{M}{M*}$ in far away points ($r\rightarrow\infty$) we
obtain a strong coincidence between different prescriptions as
\begin{eqnarray}
_{\textrm{\tiny E}}E= _{\textrm{\tiny LL}}E= _{\textrm{\tiny W}}E=
_{\textrm{\tiny P}}E= _{\textrm{\tiny M}}E&=&M.
\end{eqnarray}

\bigskip

\section{Einstein and M\o ller prescriptions' Coincidence}
From Eq.(\ref{Einstein}) and Eq.(\ref{Moller}) it is obvious that
Einstein and M\o ller expressions for the energy in GSS space-time
(\ref{GSS}) are different, generally. In this section, supposing
some conditions, we look for a suitable form for GSS metric in
which Einstein and M\o ller prescriptions give the same result.
For other forms of GSS metric Einstein and M\o ller EMPs give
different expressions for energy.

\medskip

Supposing $B(r)=A(r)^{-1}$, $F=0$ in (\ref{GSS}) for coincidence
of energy expressions of Einstein (\ref{Einstein}) and M\o ller
(\ref{Moller}) we must have

\begin{eqnarray}\label{equ}
\frac{r(1-B D -D' B r)}{2}=\frac{r^2 D B'}{2}.
\end{eqnarray}

For a special but not unusual case, i.e. $D=\digamma(r)^{1-\mu}$,
$B=\digamma(r)^\mu$, Eq.(\ref{equ}) transforms to the following
differential equation

\begin{eqnarray}
1-\digamma-r \digamma'=0.
\end{eqnarray}

Solving above differential equation, we obtain
\begin{eqnarray}\label{EM}
\digamma(r)=1+\frac{C}{r}
\end{eqnarray}
where $C$ is the integration Constant. The above equation means
that with the aforementioned conditions on GSS space-time, energy
expressions in Einstein and M\o ller prescriptions are equal only
if $B$ coefficient in line element (\ref{GSS}) has a specific
form, i.e.

\begin{eqnarray}
ds^2&=&(1+\frac{C}{r})^{\mu} dt^2-(1+\frac{C}{r})^{-\mu}
dr^2\\\nonumber &&-r^2 (d\theta^2+\sin^2(\theta)d\phi^2).
\end{eqnarray}
where for $\mu=1$ ($D=1$) we obtain $B=1+\frac{C}{r}$
(Schwarzschild like), and any other choices for $B$ do not fulfill
our purpose of Einstein and M\o ller prescriptions' coincidence
(see \cite{xulu}).

\bigskip

\section{Summary and Conclusion}
Considering Einstein, Landau-Lifshitz, Papapetrou, Weinberg, and
M\o ller EMPs, we reviewed the previous results about a 2-sphere
with radius $r$ in GSS space-time. It is obtained that these
energy expressions are different in Schwarzschild Cartesian
coordinates and coincide in Kerr-Schild Cartesian coordinates. We
used Schwarzschild Cartesian coordinates and restricted ourselves
to two special but not unusual GSS space-times and obtained
interesting results for Einstein, Landau-Lifshitz, Papapetrou,
Weinberg, and M\o ller EMPs. 

Our results can be considered as an
extension of Virbhadra's viewpoint to Schwarzschild Cartesian
coordinates which say that different EMPs may provide same bases
to define a unique quantity. Finally, we find a unique form for a
special GSS metric in which Einstein and M\o ller prescriptions
lead to the same result. For other forms of GSS space-time
Einstein and M\o ller EMPs give different expressions for the
energy. This work is one of a series of studies by the authors on energy-momentum prescriptions in general relativity~\cite{MA1, MA2, MA4}.


\begin{thebibliography}{99}


\bibitem{weinberg}Weinberg, S.: \textit{Gravitation and Cosmology: Principles and Applications of General Theory of Relativity}, (John Wiley and Sons, Inc., New York,
1972).
\bibitem{misner}Misner, C. W., Thorne, K. S. and Wheeler, J. A.: \textit{Gravitation, W. H. Freeman and Co.}, NY, p.467 and 603
(1973).
\bibitem{favata}Pinto Neto, N. and Trajtenberg P.I.: \textit{Brazilian Journal of Physics,} Vol. \textbf{30}, no. 1, Marco (2000); Favata, Marc: \textit{Phys. Rev. }\textbf{D63}
(2001).
\bibitem{moller}M\o ller, C.: \textit{Ann. Phys.} (NY) \textbf{4}, 347
(1958).
\bibitem{landau}Landau, L.D. and Lifshitz, E.M.: \textit{The Classical Theory of Fields},280 (Addison- Wesley Press,
1987).
\bibitem{bergmann}Bergmann, P. G., and Thompson R.: \textit{Phys. Rev.} \textbf{89}, 400
(1953).
\bibitem{papapetrou}Papapetrou, A.: \textit{Proc. R. Irish. Acad.} \textbf{A52}, 11
(1948).
\bibitem{tolman}Tolman, R.C.: \textit{Relativity, Thermodynamics and Cosmology}, 227 (Oxford Univ. Press,
(1934).
\bibitem{komar}Komar A.: \textit{Phys. Rev.} \textbf{113}, 934
(1959).
\bibitem{penrose}Penrose, R.: \textit{Proc. Roy. Soc. London} \textbf{A381}, 53
(1982).
\bibitem{qadir}Qadir, A. and Sharif, M.: \textit{Phys. Lett.} \textbf{A176},
331 (1992).
\bibitem{chang}Chang, C.C. Nester, J.M., Chen, C.M.: \textit{Phys. Rev. Lett.} \textbf{83}, 1897
(1999).
\bibitem{cooperstock}Cooperstock, F. I. and Sarracino, R. S.: \textit{J. Phys. A: Math. Gen.} \textbf{11}, 877
(1978).
\bibitem{bondi}Bondi, H.: \textit{Proc. R. Soc. London }\textbf{A427}, 249
(1990).
\bibitem{teleparalel}Mikhail, F. I., Wanas, M.I, Hindawi, A., and Lashin, E. I.,: \textit{ Int. J. Theor. Phys.} \textbf{32}, 1627
(1993); Vargas, T.,: \textit{ Gen. Rel. Grav.} \textbf{36}, 1255
(2004).
\bibitem{quasilocal}Penrose, R.: \textit{Proc. R. Soc. London.}
\textbf{A388}, 457 (1982); Hayward, S.A.: \textit{Phys. Rev.}
\textbf{D497}, 831 (1994); Brown, J. D., and York, Jr. J. W.:
\textit{Phys. Rev.} \textbf{D47}, 1407 (1993).
\bibitem{berq}Berqvist, G.: \textit{Class. Quantum Grav.} \textbf{9},
831(1994).
\bibitem{virbhadra1} Virbhadra, K. S.: \textit{Phys. Rev.} \textbf{D41},
1086 (1990).
\bibitem{radinschi1}Radinschi, I.: \textit{Mod. Phys. Lett.}
\textbf{A15}, 803 (2000).
\bibitem{examples}
Virbhadra, K. S.: \textit{Phys. Rev.} \textbf{D42}, 2919 (1990);
Virbhadra, K. S.: \textit{Phys. Rev.} \textbf{D42}, 1066 (1990);
Virbhadra, K. S.: \textit{Pramana} \textbf{38}, 31 (1992);
Virbhadra, K.S., and Parikh, J.C.: \textit{Phys. Lett.}
\textbf{B317}, 312 (1993); Virbhadra, K.S., and Parikh, J.C.:
\textit{Phys. Lett.} \textbf{B331}, 302 (1994); Chamorro1, A., and
Virbhadra, K. S.: \textit{Pramana} \textbf{45}, 181 (1995);
Aguirregabiria, J. M., Chamorro, A., and Virbhadra, K. S.:
\textit{Gen. Rel. Grav.} \textbf{28}, 1393 (1996); Virbhadra, K.
S.: \textit{Int. J. Mod. Phys.} \textbf{A12}, 4831 (1997); Xulu,
S. S.: \textit{Int. J. Mod. Phys.} \textbf{D7}, 773 (1998); Xulu,
S. S.: \textit{Int. J. Theor. Phys.} \textbf{37}, 1773 (1998);
Xulu, S. S.: \textit{Int. J. Mod. Phys.} \textbf{A15}, 2979
(2000); Xulu, S. S.: \textit{Int. J. Mod. Phys.} \textbf{A15},
4849 (2000);  Xulu, S. S.: \textit{Int. J. Theor. Phys.}
\textbf{39}, 1153 (2000); Xulu, S. S.: \textit{Mod. Phys. Lett.}
\textbf{A15}, 1511 (2000); Radinschi, I.: \textit{Mod. Phys.
Lett.} \textbf{A17}, 1159 (2002); Xulu, S. S.:\textit{Astrophys.
Space Sci.} \textbf{283}, 23 (2003); Vagenas, E. C.: \textit{Int.
J. Mod. Phys.} \textbf{A18}, 5781 (2003); Vagenas, E. C.:
\textit{Int. J. Mod. Phys.} \textbf{A18}, 5949 (2003); Radinschi,
I.: \textit{Chin. J. Phys.} \textbf{42}, 40 (2004); Vagenas, E.
C.:\textit{Mod. Phys. Lett.} \textbf{A19}, 213 (2004); Vagenas, E.
C.:\textit{Int. J. Mod. Phys.} \textbf{D14}, 573 (2005);
Radinschi, I.: \textit{Int. J. Mod. Phys.} \textbf{A21}, 2853
(2006); Binbay, F., Pirinccioglu, N. , Salti, M., and Aydogdu O.:
\textit{Int. J. Theor. Phys.} \textbf{46}, 2339 (2007).

\bibitem{virbb}Virbhadra, K. S.: \textit{Phys. Rev.} \textbf{D60}, 104041
(1999).
\bibitem{aguirregabiria}Aguirregabiria, J. M., Chamorro, A. and Virbhadra, K. S.: \textit{Gen. Relativ. Gravit.} \textbf{28}, 1393
(1996).
\bibitem{xulu}Xulu, S. S.: \textit{Astrophys. and Space Sc.} \textbf{283},
23 (2003).
\bibitem{palmer}Palmer, T.: \textit{Gen, Relativ. Gravit.} \textbf{12}, 149
(1980).
\bibitem{lessner}Lessner, G.: \textit{Gen. Relativ. Gravit.} \textbf{28}, 527
(1996).
\bibitem{einmol} Yang, I-Ching, and Radinschi, I.:\textit{Chin. J. Phys.} \textbf{42},
40 (2004); Matyjasek, J.: Mod. Phys. Lett. \textbf{A23}, 591
(2008).

\bibitem{vagenas1}Vagenas, E. C.:\textit{Mod. Phys.
Lett.} \textbf{A21}, 1947 (2006);

\bibitem{dym}Dymnikova, I. G.: \textit{Gen. Rel. Grav.} \textbf{24}, 235
(1992).
\bibitem{abg}A\'{y}on-Beato, E., and Garcia, A.: \textit{Phys. Lett.} \textbf{B464}, 25
(1999).
\bibitem{con}Virbhadra, K.S. and Parikh, J.C.: \textit{Phys. Lett.} \textbf{B331}, 302
(1994).
\bibitem{del}De Lorenci, V. A., Klippert, R., Novello, M., and Salim, J. M.: \textit{Phys. Lett.} \textbf{B482}, 134
(2000).
\bibitem{cha}Martinez, C., Staforelli, J. P., and Troncoso, R.: \textit{Phys. Rev.} \textbf{D74},
044028 (2006).
\bibitem{bar}Bardeen, J.: \textit{Proc.} \textbf{GR5}, Tiflis,
USSR (1968).

\bibitem{examples2} Virbhadra, K. S.: \textit{Int. J. Mod. Phys.} \textbf{A12}, 4831
(1997); Salti, M., Aydogdu, O.:\textit{Eur. Phys. J.}
\textbf{C47}, 247 (2006); Radinschi, I.:\textit{Chin. J. Phys.}
\textbf{41}, 326 (2003); Sharif, S.:\textit{Nuovo Ci.}
\textbf{B118}, 669 (2003);Radinschi, I.:\textit{Mod. Phys. Lett.}
\textbf{A16}, 673 (2001); Radinschi, I.: \textit{Horizons in World
Physics}, ed. Albert Reimer, Nova Science Publishers, Inc.,
(2005), ISBN 1-59454-320-8; Radinschi, I.:\textit{Mod. Phys.
Lett.} \textbf{A15}, 2171 (2000); Xulu, S.
S.:\textit{arXiv:gr-qc/0304081} (2003); Radinschi, I.:\textit{Rom.
Journ. Phys.} \textbf{15}, Nos. 9-10, 1223 (2005); Radinschi,
I.:\textit{FIZIKA} \textbf{B9}, 43 (2000); Salti, M., and Aydogdu,
O.:\textit{Int. J. Theor. Phys.} \textbf{45}, 2481 (2006).
\bibitem{JNW}Janis, A. I., Newman, E. T., and Winicour, J.: \textit{Phys. Rev. Lett.} \textbf{20},
878 (1968).
\bibitem{MA1}Abbassi, A., Mirshekari, S.: \textit{Int. J. Mod. Phys.} \textbf{A23}, 4569-4577 (2008)
\bibitem{MA2}Abbassi, A., Mirshekari, S.: \textit{Phys. Rev. D} \textbf{78}, 064053 (2008)
\bibitem{MA4}Mirshekari, S., Abbassi, A.: \textit{Int. J. Mod. Phys.} \textbf{A24}, 789-797 (2009)

\end{thebibliography}
\end{document}